\documentstyle[12pt]{article}
\newcounter{one}
\newcounter{two}
\newcounter{three}
\newcounter{four}
\newcounter{five}
\newcounter{six}
\begin{document}
\begin{center}
{\huge Discrete spectrum for $n$-cell potentials.}

\vspace{1.5cm}

{\Large\bf P.G. Grinevich \footnote{An essential part of this 
paper was fulfilled during the author's visit to Nantes University (France)
in February 1998. He thanks the Nantes University for hospitality.
The author was also supported by the Alexander von Humboldt-Stiftung 
(Germany), INTAS cooperation program grant No 93-166-EXT, and by the Russian 
Foundation for Fundamental Studies grant 98-01-01161.}}

\bigskip

{\bf Landau Institute for Theoretical Physics, Kosygina 2, Moscow, 
117940, Russia,\\ e-mail: pgg@landau.ac.ru}

\vspace{1cm}

{\Large\bf R.G. Novikov}

\bigskip

{\bf CNRS, UMR~6629, D\'epartement de Math\'ematiques, Universit\'e
de Nantes, BP 92208, F-44322, Nantes, cedex 03, France,\\ e-mail: 
novikov@math.univ-nantes.fr}
\end{center}

\vspace{1.5cm}

\expandafter\ifx\csname amssym.def\endcsname\relax \else\endinput\fi
%
\expandafter\edef\csname amssym.def\endcsname{%
       \catcode`\noexpand\@=\the\catcode`\@\space}
\catcode`\@=11
%

\def\undefine#1{\let#1\undefined}
\def\newsymbol#1#2#3#4#5{\let\next@\relax
 \ifnum#2=\@ne\let\next@\msafam@\else
 \ifnum#2=\tw@\let\next@\msbfam@\fi\fi
 \mathchardef#1="#3\next@#4#5}
\def\mathhexbox@#1#2#3{\relax
 \ifmmode\mathpalette{}{\m@th\mathchar"#1#2#3}%
 \else\leavevmode\hbox{$\m@th\mathchar"#1#2#3$}\fi}
\def\hexnumber@#1{\ifcase#1 0\or 1\or 2\or 3\or 4\or 5\or 6\or 7\or 8\or
 9\or A\or B\or C\or D\or E\or F\fi}

\font\tenmsa=msam10
\font\sevenmsa=msam7
\font\fivemsa=msam5
\newfam\msafam
\textfont\msafam=\tenmsa
\scriptfont\msafam=\sevenmsa
\scriptscriptfont\msafam=\fivemsa
\edef\msafam@{\hexnumber@\msafam}
\mathchardef\dabar@"0\msafam@39
\def\dashrightarrow{\mathrel{\dabar@\dabar@\mathchar"0\msafam@4B}}
\def\dashleftarrow{\mathrel{\mathchar"0\msafam@4C\dabar@\dabar@}}
\let\dasharrow\dashrightarrow
\def\ulcorner{\delimiter"4\msafam@70\msafam@70 }
\def\urcorner{\delimiter"5\msafam@71\msafam@71 }
\def\llcorner{\delimiter"4\msafam@78\msafam@78 }
\def\lrcorner{\delimiter"5\msafam@79\msafam@79 }
\def\yen{{\mathhexbox@\msafam@55}}
\def\checkmark{{\mathhexbox@\msafam@58}}
\def\circledR{{\mathhexbox@\msafam@72}}
\def\maltese{{\mathhexbox@\msafam@7A}}

\font\tenmsb=msbm10
\font\sevenmsb=msbm7
\font\fivemsb=msbm5
\newfam\msbfam
\textfont\msbfam=\tenmsb
\scriptfont\msbfam=\sevenmsb
\scriptscriptfont\msbfam=\fivemsb
\edef\msbfam@{\hexnumber@\msbfam}
\def\Bbb#1{{\fam\msbfam\relax#1}}
\def\widehat#1{\setbox\z@\hbox{$\m@th#1$}%
 \ifdim\wd\z@>\tw@ em\mathaccent"0\msbfam@5B{#1}%
 \else\mathaccent"0362{#1}\fi}
\def\widetilde#1{\setbox\z@\hbox{$\m@th#1$}%
 \ifdim\wd\z@>\tw@ em\mathaccent"0\msbfam@5D{#1}%
 \else\mathaccent"0365{#1}\fi}
\font\teneufm=eufm10
\font\seveneufm=eufm7
\font\fiveeufm=eufm5
\newfam\eufmfam
\textfont\eufmfam=\teneufm
\scriptfont\eufmfam=\seveneufm
\scriptscriptfont\eufmfam=\fiveeufm
\def\frak#1{{\fam\eufmfam\relax#1}}
\let\goth\frak

\csname amssym.def\endcsname

\expandafter\ifx\csname pre amssym.tex at\endcsname\relax \else \endinput\fi
\expandafter\chardef\csname pre amssym.tex at\endcsname=\the\catcode`\@
\catcode`\@=11
\begingroup\ifx\undefined\newsymbol \else\def\input#1 {\endgroup}\fi
\input amssym.def \relax
\newsymbol\boxdot 1200
\newsymbol\boxplus 1201
\newsymbol\boxtimes 1202
\newsymbol\square 1003
\newsymbol\blacksquare 1004
\newsymbol\centerdot 1205
\newsymbol\lozenge 1006
\newsymbol\blacklozenge 1007
\newsymbol\circlearrowright 1308
\newsymbol\circlearrowleft 1309
\undefine\rightleftharpoons
\newsymbol\rightleftharpoons 130A
\newsymbol\leftrightharpoons 130B
\newsymbol\boxminus 120C
\newsymbol\Vdash 130D
\newsymbol\Vvdash 130E
\newsymbol\vDash 130F
\newsymbol\twoheadrightarrow 1310
\newsymbol\twoheadleftarrow 1311
\newsymbol\leftleftarrows 1312
\newsymbol\rightrightarrows 1313
\newsymbol\upuparrows 1314
\newsymbol\downdownarrows 1315
\newsymbol\upharpoonright 1316
 \let\restriction\upharpoonright
\newsymbol\downharpoonright 1317
\newsymbol\upharpoonleft 1318
\newsymbol\downharpoonleft 1319
\newsymbol\rightarrowtail 131A
\newsymbol\leftarrowtail 131B
\newsymbol\leftrightarrows 131C
\newsymbol\rightleftarrows 131D
\newsymbol\Lsh 131E
\newsymbol\Rsh 131F
\newsymbol\rightsquigarrow 1320
\newsymbol\leftrightsquigarrow 1321
\newsymbol\looparrowleft 1322
\newsymbol\looparrowright 1323
\newsymbol\circeq 1324
\newsymbol\succsim 1325
\newsymbol\gtrsim 1326
\newsymbol\gtrapprox 1327
\newsymbol\multimap 1328
\newsymbol\therefore 1329
\newsymbol\because 132A
\newsymbol\doteqdot 132B
 \let\Doteq\doteqdot
\newsymbol\triangleq 132C
\newsymbol\precsim 132D
\newsymbol\lesssim 132E
\newsymbol\lessapprox 132F
\newsymbol\eqslantless 1330
\newsymbol\eqslantgtr 1331
\newsymbol\curlyeqprec 1332
\newsymbol\curlyeqsucc 1333
\newsymbol\preccurlyeq 1334
\newsymbol\leqq 1335
\newsymbol\leqslant 1336
\newsymbol\lessgtr 1337
\newsymbol\backprime 1038
\newsymbol\risingdotseq 133A
\newsymbol\fallingdotseq 133B
\newsymbol\succcurlyeq 133C
\newsymbol\geqq 133D
\newsymbol\geqslant 133E
\newsymbol\gtrless 133F
\newsymbol\sqsubset 1340
\newsymbol\sqsupset 1341
\newsymbol\vartriangleright 1342
\newsymbol\vartriangleleft 1343
\newsymbol\trianglerighteq 1344
\newsymbol\trianglelefteq 1345
\newsymbol\bigstar 1046
\newsymbol\between 1347
\newsymbol\blacktriangledown 1048
\newsymbol\blacktriangleright 1349
\newsymbol\blacktriangleleft 134A
\newsymbol\vartriangle 134D
\newsymbol\blacktriangle 104E
\newsymbol\triangledown 104F
\newsymbol\eqcirc 1350
\newsymbol\lesseqgtr 1351
\newsymbol\gtreqless 1352
\newsymbol\lesseqqgtr 1353
\newsymbol\gtreqqless 1354
\newsymbol\Rrightarrow 1356
\newsymbol\Lleftarrow 1357
\newsymbol\veebar 1259
\newsymbol\barwedge 125A
\newsymbol\doublebarwedge 125B
\undefine\angle
\newsymbol\angle 105C
\newsymbol\measuredangle 105D
\newsymbol\sphericalangle 105E
\newsymbol\varpropto 135F
\newsymbol\smallsmile 1360
\newsymbol\smallfrown 1361
\newsymbol\Subset 1362
\newsymbol\Supset 1363
\newsymbol\Cup 1264
 \let\doublecup\Cup
\newsymbol\Cap 1265
 \let\doublecap\Cap
\newsymbol\curlywedge 1266
\newsymbol\curlyvee 1267
\newsymbol\leftthreetimes 1268
\newsymbol\rightthreetimes 1269
\newsymbol\subseteqq 136A
\newsymbol\supseteqq 136B
\newsymbol\bumpeq 136C
\newsymbol\Bumpeq 136D
\newsymbol\lll 136E
 \let\llless\lll
\newsymbol\ggg 136F
 \let\gggtr\ggg
\newsymbol\circledS 1073
\newsymbol\pitchfork 1374
\newsymbol\dotplus 1275
\newsymbol\backsim 1376
\newsymbol\backsimeq 1377
\newsymbol\complement 107B
\newsymbol\intercal 127C
\newsymbol\circledcirc 127D
\newsymbol\circledast 127E
\newsymbol\circleddash 127F
\newsymbol\lvertneqq 2300
\newsymbol\gvertneqq 2301
\newsymbol\nleq 2302
\newsymbol\ngeq 2303
\newsymbol\nless 2304
\newsymbol\ngtr 2305
\newsymbol\nprec 2306
\newsymbol\nsucc 2307
\newsymbol\lneqq 2308
\newsymbol\gneqq 2309
\newsymbol\nleqslant 230A
\newsymbol\ngeqslant 230B
\newsymbol\lneq 230C
\newsymbol\gneq 230D
\newsymbol\npreceq 230E
\newsymbol\nsucceq 230F
\newsymbol\precnsim 2310
\newsymbol\succnsim 2311
\newsymbol\lnsim 2312
\newsymbol\gnsim 2313
\newsymbol\nleqq 2314
\newsymbol\ngeqq 2315
\newsymbol\precneqq 2316
\newsymbol\succneqq 2317
\newsymbol\precnapprox 2318
\newsymbol\succnapprox 2319
\newsymbol\lnapprox 231A
\newsymbol\gnapprox 231B
\newsymbol\nsim 231C
\newsymbol\ncong 231D
\newsymbol\diagup 201E
\newsymbol\diagdown 201F
\newsymbol\varsubsetneq 2320
\newsymbol\varsupsetneq 2321
\newsymbol\nsubseteqq 2322
\newsymbol\nsupseteqq 2323
\newsymbol\subsetneqq 2324
\newsymbol\supsetneqq 2325
\newsymbol\varsubsetneqq 2326
\newsymbol\varsupsetneqq 2327
\newsymbol\subsetneq 2328
\newsymbol\supsetneq 2329
\newsymbol\nsubseteq 232A
\newsymbol\nsupseteq 232B
\newsymbol\nparallel 232C
\newsymbol\nmid 232D
\newsymbol\nshortmid 232E
\newsymbol\nshortparallel 232F
\newsymbol\nvdash 2330
\newsymbol\nVdash 2331
\newsymbol\nvDash 2332
\newsymbol\nVDash 2333
\newsymbol\ntrianglerighteq 2334
\newsymbol\ntrianglelefteq 2335
\newsymbol\ntriangleleft 2336
\newsymbol\ntriangleright 2337
\newsymbol\nleftarrow 2338
\newsymbol\nrightarrow 2339
\newsymbol\nLeftarrow 233A
\newsymbol\nRightarrow 233B
\newsymbol\nLeftrightarrow 233C
\newsymbol\nleftrightarrow 233D
\newsymbol\divideontimes 223E
\newsymbol\varnothing 203F
\newsymbol\nexists 2040
\newsymbol\Finv 2060
\newsymbol\Game 2061
\newsymbol\mho 2066
\newsymbol\eth 2067
\newsymbol\eqsim 2368
\newsymbol\beth 2069
\newsymbol\gimel 206A
\newsymbol\daleth 206B
\newsymbol\lessdot 236C
\newsymbol\gtrdot 236D
\newsymbol\ltimes 226E
\newsymbol\rtimes 226F
\newsymbol\shortmid 2370
\newsymbol\shortparallel 2371
\newsymbol\smallsetminus 2272
\newsymbol\thicksim 2373
\newsymbol\thickapprox 2374
\newsymbol\approxeq 2375
\newsymbol\succapprox 2376
\newsymbol\precapprox 2377
\newsymbol\curvearrowleft 2378
\newsymbol\curvearrowright 2379
\newsymbol\digamma 207A
\newsymbol\varkappa 207B
\newsymbol\Bbbk 207C
\newsymbol\hslash 207D
\undefine\hbar
\newsymbol\hbar 207E
\newsymbol\backepsilon 237F
\catcode`\@=\csname pre amssym.tex at\endcsname

\begin{abstract} {We study the scattering problem, the Sturm-Liouville
problem and the spectral problem with periodic or skew-periodic boundary
conditions for the one-dimensional Schr\"odinger equation with an $n$-cell 
(finite periodic) potential. We give explicit upper and lower bounds for 
the distribution functions of discrete spectrum for these problems.
For the scattering problem we give, besides, explicit upper and lower
bounds for the distribution function of discrete spectrum for the case 
of potential consisting of $n$ not necessarily identical cells.  
For the scattering problem some results about transmission 
resonances are obtained.}
\end{abstract}

{\bf Mathematics Subject Classification (1991) 34L15, 34L24, 34B24.}

\section*{0. Introduction.}

We consider the one-dimensional Schr\"odinger equation
$$
-\Psi''+q_n(x)\Psi=E\Psi
\eqno(0.1)
$$
with an $n$-cell (finite-periodic) potential $q_n(x)$, i.e. 
$q_n(x)=\chi_n(x)q(x)$, where
$q(x)$ is a real-valued integrable periodic potential with period $a$, 
$\chi_n(x)$ denotes the characteristic function of the interval $[0,na]$, 
$n\in {\Bbb N}$. We study
\begin{enumerate}
\item the scattering problem on the whole line 
\item the Sturm-Liouville problem on the interval $[0,na]$, i.e. the
spectral problem on $[0,na]$ with the boundary conditions
$$
\Psi(0)\cos\alpha-\Psi'(0)\sin\alpha=0
\eqno(0.2)
$$
$$
\Psi(na)\cos\beta-\Psi'(na)\sin\beta=0, \qquad \alpha\in \Bbb R,
~~~\beta\in\Bbb R,
$$ 
and 
\item the spectral problem on $[0,na]$ with periodic $(0.3a)$ or 
skew-periodic $(0.3b)$ boundary conditions
$$
\Psi(0)=\Psi(na), \qquad \Psi'(0)=\Psi'(na),
\eqno(0.3a)
$$
$$
\Psi(0)=-\Psi(na), \qquad \Psi'(0)=-\Psi'(na),
\eqno(0.3b)
$$
\end{enumerate}
We discuss relations between spectral data for these problems and spectral
data for the one-dimensional Schr\"odinger equation on the whole line with 
the related periodic potential $q$.

In the present paper we obtain, in particular, the following estimates:
\begin{itemize}
\item
if $F_{sc}^{(n)}(\Omega)$ is the distribution function of discrete 
spectrum for the scattering problem for $(0.1)$ on the whole line 
(the number of eigenvalues in $\Omega \subset ]-\infty,0]$ for 
this problem), then
$$
\Bigl|F_{sc}^{(n)}(]-\infty,E[)-[\pi^{-1}nap(E)]\Bigr|\le 1 \qquad\mbox
{for    }E\le 0, 
\eqno(0.4)
$$
\item
if $F^{(n)}(\Omega)$ is the distribution function of discrete spectrum 
for $(0.1)$  on $[0,na]$ with $(0.2)$ (the number of eigenvalues in 
$\Omega\subset\Bbb R$ for this problem), then
$$
\Bigl|F^{(n)}(]-\infty,E])-[\pi^{-1}nap(E)]\Bigr|\le 1 \qquad\mbox
{for    }E\in\Bbb R,\ 0\le\alpha\le\beta\le\pi,\ \beta\ne 0,\ \alpha\ne\pi, 
\eqno(0.5a)
$$
$$
\Bigl|F^{(n)}(]-\infty,E[)-[\pi^{-1}nap(E)]-1\Bigr|\le 1 \qquad\mbox
{for    }E\in\Bbb R,\ \ 0<\beta<\alpha<\pi, 
\eqno(0.5b)
$$
\item
if $F^{(n)}(\Omega)$ is the distribution function of discrete spectrum for
$(0.1)$ on $[0,na]$ with $(0.3a)$ or $(0.3b)$ (the sum of multiciplities
of eifenvalues in $\Omega\subset\Bbb R$	for this problem), then
$$
[\pi^{-1}nap(E)]\le F^{(n)}(]-\infty,E[) \le [\pi^{-1}nap(E)]+1	   
\qquad\mbox{for    }E\in\Bbb R,
\eqno(0.6)
$$
where $p(E)$ is the real part of the global quasimomentum for the 
related periodic potential $q(x), [r]$ is the integer part of $r\ge 0$.
\end{itemize}

To obtain $(0.4), (0.5)$ we use, in particular, the technique presented
in Chapter 8 of \cite{CL} and some arguments of \cite{JM}. The estimate
$(0.6)$ follows, actually, from well-known results presented in Chapter 21 
of \cite{T}.

The estimates $(0.4) - (0.6)$ and additional estimates for the distribution 
functions of discrete spectrum are given in Theorems 1, 2, Corollaries 1, 2, 
and by the formulas $(2.22), (2.23), (2.29)-(2.32)$ in Section 2 of the 
present paper.

As a corollary of $(0.5), (0.6)$ one can obtain the following formula of
\cite{Sh}:
$$
\lim_{n\to\infty}(na)^{-1} F^{(n)}(]-\infty,E])=\pi^{-1}p(E),
\quad E\in\Bbb R,
\eqno(0.7)
$$
where $F^{(n)}(\Omega)$ is the distribution function for $(0.1)$ on
$[0,na]$ with $(0.2)$ or $(0.3a)$ or $(0.3b)$, $p(E)$ is the real part 
of the global 
quasimomentum for the related periodic potential. The formula $(0.7)$
for the case of smooth potential is a particular case of results of
\cite{Sh} about density of states of multidimensional selfadjoint
elliptic operators with almost periodic coefficients. The formula $(0.7)$
(for the case of continuous potential) follows from result of \cite{JM}
about density of states for the one-dimensional Schr\"odinger equation 
with almost periodic potential. Note that the methods of \cite{Sh} 
and \cite{JM} are very different.

Remark. Less precise estimates instead of (0.4) and (0.5) follow
directly from (0.6) and well-known results (see Theorems 1.1, 2.1, 3.1 
of Chapter 8 of \cite{CL} and, for example, $\S 1$ of Chapter 1 of 
\cite{NMPZ}) about zeros of eigenfunctions of the one-dimensional
Schr\"odinger operator. Probably, one can generalize such an approach to 
the multidimensional case. Concerning the distribution function of discrete
spectrum for the multidimensional Schr\"odinger operator with a finite
periodic potential with periodic boundary conditions see the proof of
Theorem XIII.101 of \cite{RS}. Concerning results about zeros of
eigenfunctions of multidimensional Schr\"odinger operator see \cite{Ku} and
subsequent references given there and also $\S 6$ of Chapter \Roman{six} 
of \cite{CH}.

The transmission resonances for the scattering problem for $(0.1)$ 
on the whole line are also considered in the present paper.  
An energy $E$ is a transmission resonance iff $E\in\Bbb R_+$ and the 
reflection coefficients are equal to zero at this energy. The main 
features of the transmission resonances for an $n$-cell scatterer 
were discussed in \cite{SWM}, \cite{RRT}. 
In the present paper (Proposition 1, the 
formula $(2.25)$) we give the following additional results about the 
transmission resonances: if $E\in\Bbb R_+$ is a double eigenvalue for 
$(0.1)$	on $[0,na]$ with $(0.3a)$ or $(0.3b)$, then $E$ is a transmission 
resonances, and all $n$-dependent transmission resonances have this origin; 
there are no transmission resonances in the forbidden energy set for the 
the related periodic potential; if $q(x)\not\equiv 0$, then
$$
(na)^{-1}\Phi_{sc}^{(n)}(]0,E])-\pi^{-1}(p(E)-p(0))= O(n^{-1})
$$
as $n\to\infty$, where $\Phi_{sc}^{(n)}(\Omega)$ 
is the number of transmission resonances in $\Omega\subset\Bbb R_+$ 
for an $n$-cell scatterer, $p(E)$ is the real part of the global 
quasimomentum for related periodic potential $q(x)$.

We consider also the one-dimensional Schr\"odinger equation
$$ 
-\psi^{\prime\prime} + q(x)\psi = E\psi,\ \ x\in{\Bbb R},
\eqno(0.8)
$$
with a potential consisting of $n$ not necessarily identical cells. More
precisely, we suppose that : ${\Bbb R}=\cup_{j=1}^nI_j$, 
where $I_1=] -\infty, x_1]$, $I_j=[x_{j-1},x_j]$ for 
$1<j<n$, $I_n=[x_{n-1},+\infty [$,
$-\infty < x_{j-1}<x_j< +\infty$ for $1<j<n$; $q(x)=\sum_{j=1}^nq_j(x)$,
where $q_j\in L^1({\Bbb R})$, $q_j={\bar q}_j$, $supp\,q_j\subseteq I_j$
for $1\le j\le n$ and, in addition, $(1+|x|)q_1(x)$ and
$(1+|x|)q_n(x)$ are also integrable on ${\Bbb R}$.

In the present paper (Theorem 3) we obtain, in particular, the following
estimate
$$|F(] -\infty, E[)-\sum_{j=1}^nF_j(] -\infty, E[)|\le n-1\ \ {\rm for}\ \
E\le 0, \eqno(0.9)$$
where $F(] -\infty, E[)$, $(F_j(] -\infty, E[)$, resp.) denotes the
distribution function of discrete spectrum for the scattering problem
for (0.8) (for the one-dimensional Schr\"odinger equation with the 
potential $q_j$, resp.) on the whole line.

In addition, for $E=0$ we have the estimate (2.36) obtained earlier in
\cite{AKM2} as a development of results of \cite{K} and \cite{SV}.

Additional indications conserning preceeding works are given in Section 2 of 
the present paper. In connection with results discussed in the present 
paper it is useful to see also the review given in \S 17 of
\cite{RSS} and \cite{KS} and the results given in \cite{ZV}.
\section*{1. Definitions, notations, assumptions and some known facts.}

We consider the one-dimensional Schr\"odinger equation
$$
-\frac{d^2}{dx^2}\Psi+q_n(x)\Psi=E\Psi, \quad x\in\Bbb R,
\eqno(1.1)
$$
where $q_n(x)$ is an $n$-cell potential, i.e.
$$
q_n(x)=\sum_{j=0}^{n-1} q_1(x-ja), \qquad a\in\Bbb R_+ ,
\eqno(1.2)
$$
$$
q_1\in L^1(\Bbb R), \qquad q_1=\bar q_1,\qquad\hbox{supp }q_1\in[0,a] .
\eqno(1.3)
$$

{\bf First,} we consider the scattering problem for the equation $(1.1)$ 
on the
whole line: we consider wave functions describing scattering with 
incident waves for positive energies and bound states for negative
energies. We recall some definitions and facts of the scattering theory
for the Schr\"odinger equation
$$
-\Psi''+v(x)\Psi=E\Psi
\eqno(1.4)
$$
where
$$
v\in L^1(\Bbb R), \quad v=\bar v, \qquad \int\limits_{\Bbb R}
(1+|x|)|v(x)|dx<\infty
\eqno(1.5)
$$
(see, for example, \cite{F}). Let an incident wave be described by
$e^{ikx}, k\in\Bbb R, k^2=E>0$. Then the scattering is described by the
wave function $\Psi^+(x,k)$ defined as a solution of $(1.4)$ such that
$$
\Psi^+(x,k)=e^{ikx}-\frac{\pi i}{|k|}e^{i|k||x|}f(k,|k|\frac{x}{|x|})+
{\it o}(1)\quad\hbox{as }x\to\infty
\eqno(1.6)
$$
for some $f(k,{\it l}), {\it l}\in\Bbb R, {\it l}^2=k^2$, which is the 
scattering amplitude. The following formulas connect the scattering amplitude
$f$ and the scatering matrix ${\it S}(k)=
({\it s}_{ij}(k)), k\in\Bbb R_+$:
$$
\begin{array}{l}
{\it s}_{11}(k)=1-\pi ik^{-1}f(-k,-k), \quad {\it s}_{12}(k)=
-\pi ik^{-1}f(-k,k), \\
\mathstrut\\
{\it s}_{21}(k)=-\pi ik^{-1}f(k,-k), \quad {\it s}_{22}(k)=
1-\pi ik^{-1}f(k,k).
\end{array}
\eqno(1.7)
$$

The bound states energies $E_j$ are defined as the discrete spectrum 
and the bound states $\Psi_j(x)$ are defined as related eigenfunctions
of the spectral problem $(1.4)$ in $L^2(\Bbb R)$. We recall that, under
assumption $(1.5)$, 
$$
S(k),\ \ k\in{\Bbb R}_+, \ \hbox{ is unitary and }\ \ 
{\it s}_{11}(k)={\it s}_{22}(k),
\eqno(1.8)
$$
each eigenvalue $E_j$ is negative and simple and the total number $m$ 
of these eigenvalues is finite
$$
E_1<E_2<\ldots<E_m<0,\ \ \ m<\infty.
\eqno(1.9)
$$

For $v=q_n$ we will write $\Psi^+, f, S, {\it s}_{ij}, E_j, \Psi_j$
as $\Psi_n^+, f_n, S_n, {\it s}_{ij}^{(n)}, E_j^{(n)}, \Psi_j^{(n)}$.
						     
{\bf Second,} we consider the spectral problem $(1.1)$ on the interval 
$[0,na]$ with the boundary conditions
$$
\Psi(0)\cos\alpha-\Psi'(0)\sin\alpha=0
$$
$$
\Psi(na)\cos\beta-\Psi'(na)\sin\beta=0, \quad\alpha\in\Bbb R,
\quad\beta\in\Bbb R.
\eqno(1.10)
$$
Without loss of generality we may assume that
$$
0\le\alpha <\pi, \quad 0<\beta\le\pi.
\eqno(1.11)
$$

{\bf Third,} we consider the spectral problem $(1.1)$ on the 
interval $[0,na]$ with the boundary conditions
$$
\Psi(0)=\Psi(na), \quad\Psi'(0)=\Psi'(na)
\eqno(1.12a)
$$
or with the boundary conditions
$$
\Psi(0)=-\Psi(na), \quad\Psi'(0)=-\Psi'(na).
\eqno(1.12b)
$$

On the other hand, we consider the one-dimensional Schr\"odinger 
equation
$$
-\frac{d^2}{dx^2}\Psi+q(x)\Psi=E\Psi, \quad x\in\Bbb R,
\eqno(1.13)
$$
where $q$ is the following periodic potential
$$
q(x)=\sum_{j=-\infty}^{\infty} q_1(x-ja).
\eqno(1.14)
$$

We recall some definitions and facts of spectral theory for the equation
$(1.13)$ on the whole line (see, for example, \S 17 of \cite{RSS} and 
Chapter \Roman{two} of \cite{NMPZ}).

The monodromy operator $M(E)$ is defined as the translation operator by 
the period $a$ in the two-dimensional space of solutions of $(1.13)$
at fixed $E$. If a basis in this space is fixed, then one can consider
$M(E)$ as a $2\times 2$ matrix. For all $E$  $det M(E)=1$. 
The eigenvalues of $M(E)$ are of the form
$$
\lambda_1=1/\lambda_2=\left(Tr\;M(E)+\sqrt{\bigl( (Tr\;M(E)\bigr)^2-4}
\right)/2=e^{i\varphi (E)},       
\eqno(1.15)
$$
where
$$
2\cos{\varphi(E)}=Tr\;M(E).
\eqno(1.16)
$$

The Bloch solutions are defined as eigenvectors of $M(E)$.

The allowed and forbidden Bloch zones are defined by the formulas
$$
\bigcup_{j\in\ J}\Lambda_j^a=\Lambda^a=\left\{ E\in\Bbb R
\bigl|\bigr.\left| Tr\;M(E)\right|\le 2\right\} \qquad\mbox{allowed zones}
\eqno(1.17)
$$
$$
\bigcup_{j\in\ J}\Lambda_j^f=\Lambda^f=\left\{ E\in\Bbb R
\bigl|\bigr.\left| Tr\;M(E)\right| > 2\right\} \qquad\mbox{forbidden zones}
\eqno(1.18)
$$
where either $J=\Bbb N$	or $J=\{ 1,\dots ,m\},\,\,m\in\Bbb N;\;\;\Lambda_j^a,
\Lambda^f$ are connected intervals (closed for the case $(1.17)$ and open
for the case $(1.18)$) such that
$$
\sup \limits_{E\in\Lambda_j^a} E < \inf\limits_{E\in\Lambda_{j+1}^a} E,\quad
\sup\limits_{E\in\Lambda_j^f} E < \inf\limits_{E\in\Lambda_{j+1}^f} E, 
\quad\mbox{for}\quad j,\, j+1\in J,
\eqno(1.19)
$$
in addition, $\Lambda_1^f=]-\infty,\lambda_0 [$.

The real part of the global quasimomentum $p(E)$ is 
defined as a real-valued continuous 
nondecreasing function such that: $p(E)$ is constant in each forbidden
zone $\Lambda_j^f, \, p(E)=0$ for $E\in\Lambda_1^f$, the phase $\varphi(E)
=ap(E)$ is a solution of $(1.16)$ for each allowed zone $\Lambda_j^a$.
Note that 
$$
\pi^{-1}ap(E)={\it l}_j\in\Bbb N\cup 0\ \hbox{ for } E\in\bar\Lambda_j^f
\eqno(1.20)
$$
(the closure of $\Lambda_j^f$) for each $j\in J,\,{\it l}_1=0,\,
{\it l}_j <{\it l}_{j+1}$ for $j,j+1\in J$.

\section*{2. The main new results and the preceeding results.}

In the present paper we discuss relations between spectral data 
for $(1.1)$  for the first, the second or the third case 
described above and spectral data for $(1.13)$.

To start with, we discuss some results of \cite{SWM}, \cite{RRT}, 
\cite{R}, \cite{SV}. In \cite{SWM} the following formulas are given, 
in particular:
$$
\frac{R_n}{T_n}=\frac{\displaystyle\sin{n\varphi}}
{\displaystyle\sin{\varphi}}
\,\,\frac{R_1}{T_1}
\eqno(2.1)
$$ 
$$
\begin{array}{l}
\frac{1}{T_n}=\frac{\displaystyle 1}{\displaystyle\sin{\varphi}}
\left(\frac{1}{T_1}\,\sin{n\varphi}-
\sin{(n-1)\varphi}\right),\\
\mathstrut\\
R_n=R_1\left( 1-\frac{\displaystyle\sin{(n-1)\varphi}}
{\displaystyle\sin{\varphi}}T_1\right)^{-1},
\end{array}
\eqno(2.2)
$$
\medskip
$$
M(E)=\left(\begin{array}{rr}
\frac{\displaystyle 1}{\displaystyle{T_1}}\;&
-\frac{\displaystyle\bar{R_1}}{\displaystyle{T_1}}\\
\mathstrut & \mathstrut \\
-\frac{\displaystyle{R_1}}{\displaystyle{T_1}\;}&
\frac{\displaystyle 1}{\displaystyle{T_1}}
\end{array}\right)
\eqno(2.3)
$$

\bigskip
\noindent
in the basis of solutions $\psi_{\pm}(x,k)$ such that $\psi_{\pm}(0,k)
=1$, $\psi_{\pm}^{\prime}(0,k)=\pm ik$,
$$
\cos\varphi = Re(1/T_1),
\eqno(2.4)
$$
where $T_n={\it s}_{22}^{(n)}(E) e^{ikna},\;R_n={\it s}_{21}^{(n)}(E),\;
\varphi=\varphi(E)$ is the Bloch phase from $(1.15),\,(1.16),\;
E=k^2,\;k\in\Bbb R_+$.

The formulas $(2.1)$--$(2.4)$ (taking into account $(1.8)$) describe 
relations between the scattering matrix ${\it s}_{ij}^{(n)}(k),\;
k\in\Bbb R_+$, for $(1.1)$ and spectral data for $(1.13)$ in a very
complete way. A proper discussion is given in \cite{SWM}. Some similar 
results are given also in \cite{RRT}. Conserning more old results 
in this direction see \cite{R} and references given in \cite{SWM}, 
\cite{RRT}. The discrete spectrum for the scattering problem for $(1.1)$ 
on the whole line was discussed	in \cite{R}, \cite{SV}. The paper 
\cite{R} deals with the particular case when 
$q_1(\frac{a}{2}+x)=q_1(\frac{a}{2}-x)$ and results of \cite{R} conserning 
the discrete spectrum imply a lower bound for the distribution function 
of discrete spectrum $F_{sc}^{(n)}(\Sigma)$ for this case. In \cite{SV}
the total number of bound 
states for an $n$-cell scatterer $q_n$ is given in terms of certain
quantities characterizing the single scatterer $q_1$. However, 
in \cite{SV} the distribution function	      
$$
F_{sc}^{(n)}(\Sigma)=\;\#\{E_{j}^{(n)}\in\Sigma\}
\eqno(2.5)
$$
(the number of bound states with energies in an interval $\Sigma\subset]
-\infty,\,0[$) is not considered for $\Sigma\ne]-\infty,\,0[$ and
manifestations of the Bloch zone structure for $E_{j}^{(n)}$ are
not discussed.

In the present paper we obtain, in particular, the following result.

{\bf Theorem 1.}{\it Under assumptions $(1.2),\;(1.3),\;(1.14)$, the 
following formulas hold:
$$
\left| F_{sc}^{(n)}(]-\infty,\,E[)-\Bigl[\frac{nap(E)}{\pi}\Bigr]\right|\le 1
\quad\mbox{for  }E\in ]-\infty,\,0]\;,
\eqno(2.6)
$$
$$
\left[\frac{nap(E)}{\pi}\right]\le F_{sc}^{(n)}(]-\infty,\,E[) 
\le\left[\frac{nap(E)}
{\pi}\right]+1\quad\mbox{for  }E\in]-\infty,\;0]\setminus\bar\Lambda^f,
\eqno(2.7)
$$
where $F_{sc}^{(n)}(\Sigma)$ is the distribution function of discrete 
spectrum for the scattering problem $(1.1),\;\; p(E)$ is the real part 
of the global 
quasimomentum and $\bar\Lambda^f$ is the closure of the forbidden
energy set for the spectral problem $(1.13),\;[r]$ is the integer part 
of $r\ge 0$.}

The proof of {\bf Theorem 1} is given in Section 4.

Using $(2.6)$ we obtain the following corollary.

{\bf Corollary 1.} {\it Under assumptions $(1.2),\,(1.3),\,(1.14)$, the
following formulas hold:
$$
\begin{array}{l}
F_{sc}^{(n)}(\bar\Lambda^f_{j}\bigcap ]-\infty,\,0[) \le 2
\quad\mbox{for  }j\in J,\\
F_{sc}^{(n)}(\bar\Lambda^f_{1}\bigcap ]-\infty,\,0[) \le 1 ,
\end{array}
\eqno(2.8)
$$
where $\bar\Lambda_j^f$ is the closure of the forbidden zone $\Lambda_j^f$
for $(1.13)$.}

The proof of {Corollary 1} is given in Section 4.

Consider now the eigenvalues $E_j^{(n)}$ and the distribution function
$$
F^{(n)}(\Sigma)=\;\# \{E_{j}^{(n)}\in\Sigma\}
\eqno(2.9)
$$
(the number of eigenvalues in an interval $\Sigma\subset\Bbb R$) for the
spectral problem $(1.1),\;(1.10)$.

{\bf Theorem 2.}{\it Under assumptions $(1.2),\;(1.3),\;(1.11),\;(1.14)$,
the following formulas hold:
$$
\left[\frac{nap(E)}{\pi}\right]-1\le F^{(n)}(]-\infty,\,E[) \le\left[
\frac{nap(E)}{\pi}\right]\quad\mbox{for  }E\in\Bbb R ,\;\alpha =0,\;\;
\beta =\pi,
\eqno(2.10a)
$$
$$
F^{(n)}(]-\infty,\,E[)=\Bigl[\frac{nap(E)}{\pi}\Bigr]\quad\mbox{for  }
E\in\Bbb R \setminus\bar\Lambda^f,\;\;\alpha =0,\;\;\beta =\pi,
\eqno(2.10b)
$$
$$
\left| F^{(n)}(]-\infty,\,E[)-\Bigl[\frac{nap(E)}{\pi}\Bigr]\right|\le 1
\quad\mbox{for  }E\in\Bbb R,\;\;\alpha <\beta,
\eqno(2.11a)
$$
$$
\left[\frac{nap(E)}{\pi}\right]\le F^{(n)}(]-\infty,\,E[) \le\left[
\frac{nap(E)}{\pi}\right]+1\quad\mbox{for  }E\in\Bbb R \setminus\bar\Lambda^f,
\;\alpha <\beta,
\eqno(2.11b)
$$
$$
\left| F^{(n)}(]-\infty,\,E[)-\Bigl[\frac{nap(E)}{\pi}\Bigr] -1\right|\le 1
\quad\mbox{for  }E\in\Bbb R,\;\;\beta <\alpha ,
\eqno(2.12a)
$$
$$
\left[\frac{nap(E)}{\pi}\right] +1\le F^{(n)}(]-\infty,\,E[) \le\left[
\frac{nap(E)}{\pi}\right]+2\quad\mbox{for  }E\in\Bbb R \setminus\bar\Lambda^f,
\;\beta <\alpha,
\eqno(2.12b)
$$
$$
\left[\frac{nap(E)}{\pi}\right]\le F^{(n)}(]-\infty,\,E[) \le\left[
\frac{nap(E)}{\pi}\right]+1\quad\mbox{for  }E\in\Bbb R ,\;\alpha =\beta,
\eqno(2.13a)
$$
$$
F^{(n)}(]-\infty,\,E[)=\Bigl[\frac{nap(E)}{\pi}\Bigr] +1\quad\mbox{for  }
E\in\Bbb R \setminus\bar\Lambda^f,\;\;\alpha =\beta ,
\eqno(2.13b)
$$
where $F^{(n)}(\Sigma)$ is the distribution function for the spectral
problem $(1.1),\;(1.10),\;p(E)$ is the real part of the 
global quasimomentum and 
$\bar\Lambda^f$ is the closure of the forbidden
energy set for the spectral problem $(1.13),\;[r]$ is the integer part 
of $r\ge 0$.}

The proof of {\bf Theorem 2} is given in Section 4.

Using $(2.10)$--$(2.13)$ we obtain the following corollary.

{\bf Corollary 2.} {\it Under assumptions $(1.2),\,(1.3),\,(1.11),\,
(1.14)$, the following formulas hold:
$$
F^{(n)}(\bar\Lambda_1^f)=0\quad\mbox{for  }\alpha =0,\;\;\beta =\pi,
\eqno(2.14a)
$$
$$
F^{(n)}(\bar\Lambda_j^f)=1\quad\mbox{for  }j\in J\setminus 1,\;
\alpha =0,\;\;\beta =\pi,
\eqno(2.14b)
$$
$$
F^{(n)}(\bar\Lambda_j^f)\le 2\quad\mbox{for  }j\in J,\;\alpha <\beta ,
\eqno(2.15a)
$$
$$
F^{(n)}(\bar\Lambda_1^f)\le 1\quad\mbox{for  }\alpha <\beta ,
\eqno(2.15b)
$$
$$
F^{(n)}(\bar\Lambda_j^f)\le 2\quad\mbox{for  }j\in J,\;\beta <\alpha,
\eqno(2.16a)
$$
$$
F^{(n)}(\bar\Lambda_j^f)=1\quad\mbox{for  }j\in J,\;\alpha =\beta ,
\eqno(2.16b)
$$
where $\bar\Lambda_j^f$ is the closure of the forbidden zone $\Lambda_j$
for $(1.13)$.}

The proof of the {\bf Corollary 2} is given in Section 4.

Consider now the eigenvalues $E_j^{(n)}$ for $(1.1)$ with $(1.12a)$,
the eigenvalues $\tilde E_j^{(n)}$ for $(1.1)$ with $(1.12b)$, and
the related distribution functions
$$
F^{(n)}(\Omega)=\sum_{E_j^{(n)}\in\,\Omega} m(E_j^{(n)}),\qquad
\tilde F^{(n)}(\Omega)=\sum_{\tilde E_j^{(n)}\in\,\Omega} 
m(\tilde E_j^{(n)}),
\eqno(2.17)
$$
where $\Omega$ is a subset of $\Bbb R$, $\;m(E_j^{(n)}),\;
m(\tilde E_j^{(n)})\in\,\{1,2\}$ are the multiplicities of $E_j^{(n)},\;
\tilde E_j^{(n)}$.

Under assumptions $(1.2),\,(1.3),\,(1.14)$, the following statements
are valid:
$$
\begin{array}{l}
\hbox{\it a number E is a simple eigenvalue for (1.1) with (1.12a) iff}\\
\mathstrut\\
(2\pi )^{-1}nap(E)\in{\Bbb N}\cup 0,\quad E\in\bar\Lambda^f
\setminus\Lambda^f,
\end{array}
\eqno(2.18)
$$
\\
$$
\begin{array}{l}
\hbox{a number E is a double eigenvalue for (1.1) with (1.12a) iff}\\
\mathstrut\\
(2\pi )^{-1}nap(E)\in{\Bbb N}\cup 0,\quad E\in{\Bbb R}\setminus\bar
\Lambda^f,
\end{array}
\eqno(2.19)
$$
\\
$$
\begin{array}{l}
\hbox{a number E is a simple eigenvalue for (1.1) with (1.12b) iff}\\
\mathstrut\\
(2\pi )^{-1}(nap(E)-\pi )\in{\Bbb N}\cup 0,\quad E\in\bar\Lambda^f
\setminus\Lambda^f,
\end{array}
\eqno(2.20)
$$
\\
$$
\begin{array}{l}
\hbox{a number E is a double eigenvalue for (1.1) with (1.12b) iff}\\
\mathstrut\\
(2\pi )^{-1}(nap(E)-\pi )\in{\Bbb N}\cup 0,\quad E\in{\Bbb R}
\setminus\bar\Lambda^f,
\end{array}
\eqno(2.21)
$$
if $F^{(n)}(\Omega)$ is the distribution function for $(1.1)$ with 
$(1.12a)$, then
$$
\begin{array}{l}
[(2\pi )^{-1}nap(E)]\le F^{(n)}(]-\infty ,E])\le [(2\pi )^{-1}nap(E)]+1,\\
\mathstrut\\
F^{(n)}(\Lambda^f)=0,
\end{array}
\eqno(2.22)
$$
if $\tilde F^{(n)}(\Omega)$ is the distribution function for $(1.1)$ with 
$(1.12b)$, then
$$
\begin{array}{l}
[(2\pi )^{-1}nap(E)]\le \tilde F^{(n)}(]-\infty ,E])\le [(2\pi )^{-1}
nap(E)]+1,\\
\mathstrut\\
\tilde F^{(n)}(\Lambda^f)=0,
\end{array}
\eqno(2.23)
$$
where $p(E)$ is the real part of the global quasimomentum and 
$\bar\Lambda^f$ is the 
closure of the forbidden energy set $\Lambda^f$ for $(1.13),\;[r]$ 
is the integer part of $r\ge 0$.

Using known properties of $p(E)$ and $E_j=E_j^{(1)},\;\tilde E_j=
\tilde E_j^{(1)}$ (see, for example, \cite{RSS}, \S 17 and \cite{CL},
 Chapter 8), we obtain these statements, first, for $n=1$. 
Then we reduce the general case to the case $n=1$ considering 
$q$ from $(1.14)$ as a potential with period $na$. One can obtain 
also these statements using well-known results presented in 
Chapter 21 of \cite{T} and the 
definitions of $p(E)$ and $F^{(n)}(\Omega),\;\tilde F^{(n)}(\Omega) $.

Consider now the points of perfect transmission (transmission resonances)
for the scattering problem $(1.1)$, i.e. the points $\lambda_j^{(n)}
\in\Bbb R_+$ such that $|{\it s}_{ii}^{(n)}(\lambda_j^{(n)})|=1$.

In \cite{SWM} it is shown (using (2.1)) that, for $E\in{\Bbb R}_+$,
$$
E \ \ \ \hbox{is a point of perfect transmission, i.e.}  
\ \ \ |{\it s}_{ii}^{(n)}(E)|=1,
$$
$$
\hbox{\bf{either}  \ \ if \ }|{\it s}_{ii}^{(1)}(E)|=1
\hbox{\ \ \ \ {\bf or \ \ \ \ if} \ }
\sin{n\varphi (E)}=0,\;\sin\varphi (E)\ne 0,
\eqno(2.24)
$$
where $\varphi (E)$ is defined by $(1.16)$. The same result is given also
in \cite{RRT}.

In the present paper in connection with transmission resonances we
obtain the following result.

{\bf Proposition 1.}{\it Under assumptions $(1.2),\,(1.3),\,(1.14)$, 
the following statements are valid:
$$
\begin{array}{l}
\hbox{\it if}\ E\in\Bbb R_+\ \hbox{\it is a double eigenvalue for}\ 
(1.1)\ \hbox{\it with}\ (1.12a)\ \hbox{\it or with}\ (1.12b),\\
\hbox{\it then}\ |{\it s}_{ii}^{(n)}(E)|=1,
\end{array}
\eqno(2.25)
$$

$$
\begin{array}{l}
\hbox{\it if}\  \sin{n\varphi (E)}=0,\;\sin\varphi (E)\ne 0,\\
\hbox{\it then}\ E\ \hbox{\it }\ \hbox{\it is a double eigenvalue for}\ 
(1.1)\ \hbox{\it  with}\ (1.12a)\ \hbox{\it or with}\ (1.12b),
\hphantom{a}
\end{array} 
\eqno(2.26)
$$

$$
\begin{array}{l}
\hbox{\it if}\ |{\it s}_{ii}^{(n)}(E)|=1,\;\,E\in\Bbb R_+,\\
\hbox{\it then}\ E\in\Lambda^a,	
\hphantom{aaaaaaaaaaaaaaaaaaaaaaaaaaaaaaaaaaaaaaaaaaaaaa}
\end{array}
\eqno(2.27)
$$
\\
where $s_{ii}^{(n)}(E)$ is the transmission coefficient for $(1.1)$,
$\varphi (E)$ is the Bloch phase and $\Lambda^a$ is the allowed energy 
set for $(1.13)$.}

To prove the statement $(2.25)$ we calculate ${\it s}_{ii}^{(n)}(E)$ using
that in this case each solution of $(1.1)$ on $[0,\, na]$ satisfies
$(1.12a)$ or $(1.12b)$. The statement $(2.26)$ follows from $(2.19)$,
$(2.21)$ and properties of $\varphi (E)$. The statement $(2.27)$ follows 
for example, from $(2.24)$, $(2.4)$ and properties of $\varphi (E)$. 

For $q_1(x)\not\equiv 0$ we consider also the distribution function 
of transmissions resonances
$$
\Phi_{sc}^{(n)}(\Sigma )=\;\#\{\lambda_j^{(n)}\in\Sigma\}\;\;
(\mbox{the number of transmission resonances in  }\Sigma\subset\Bbb R_+).
\eqno(2.28)
$$

Under assumptions $(1.2),\,(1.3),\,(1.14)$, as a corollary of 
{\bf Theorems 1, 2} the statements $(2.18)$--$(2.24)$ and 
{\bf Proposition 1} we obtain the following statements:

{\it if $F_{sc}^{(n)}(\Sigma)$ is the distribution function of discrete
spectrum for the scattering problem $(1.1)$, then, for $E\ge 0$,
$$
\begin{array}{l}
\lim\limits_{n\to\infty}(na)^{-1}F_{sc}^{(n)}(]-\infty,E[)=\pi^{-1}p(E),\\
\mathstrut\\
(na)^{-1}F_{sc}^{(n)}(]-\infty,E[)-(na)^{-1}\le \pi^{-1}p(E)\le 
(na)^{-1}F_{sc}^{(n)}(]-\infty,E[)+2(na)^{-1},
\end{array}
\eqno(2.29)
$$

if $\Phi_{sc}^{(n)}(\Sigma)$ is the distribution function of transmission
resonances for the scattering problem $(1.1)$, $q_1(x)\not\equiv 0$, 
then, for $E\ge 0$,
$$
\begin{array}{l}
\lim\limits_{n\to\infty}(na)^{-1}\Phi_{sc}^{(n)}(]0,E])=
\pi^{-1}(p(E)-p(0)),\\
\mathstrut\\
(na)^{-1}\Phi_{sc}^{(n)}(]0,E])-\pi^{-1}(p(E)-p(0))={\it O}(n^{-1}), 
\;\mbox{as  }n\to\infty,
\end{array}
\eqno(2.30)
$$
if $F^{(n)}(\Sigma)$ is the distribution function for $(1.1)$, with 
$(1.10)$, then, for $E\in\Bbb R$,
$$
\lim_{n\to\infty}(na)^{-1}F^{(n)}(]-\infty,E])=\pi^{-1}p(E),\\
\eqno(2.31a)
$$
$$
(na)^{-1}F^{(n)}(]-\infty,E])-(na)^{-1}\le \pi^{-1}p(E)\le
$$
$$
\le (na)^{-1}F^{(n)}(]-\infty,E])+2(na)^{-1},
\eqno(2.31b)
$$
$$
\quad\mbox{for  }0\le\alpha\le\beta\le\pi ,\;0<\beta ,\;\alpha <\pi ,
$$
\smallskip
$$
(na)^{-1}F^{(n)}(]-\infty,E]) \le \pi^{-1}p(E)\le
$$
$$
\le (na)^{-1}F^{(n)}(]-\infty,E])+3(na)^{-1},
\eqno(2.31c)
$$
$$
\quad\mbox{for  }0<\beta <\alpha <\pi ,
$$
if $F^{(n)}(\Sigma)$ is the distribution function for $(1.1)$, with 
$(1.12a)$ or $(1.12b)$ then, for $E\in\Bbb R$,
$$
\lim_{n\to\infty}(na)^{-1}F^{(n)}(]-\infty,E])=\pi^{-1}p(E),
\eqno(2.32a)
$$
$$
(na)^{-1}F^{(n)}(]-\infty,E])-(na)^{-1}\le \pi^{-1}p(E)\le
$$
$$
\le (na)^{-1}F^{(n)}(]-\infty,E])+(na)^{-1},
\eqno(2.32b)
$$
where $p(E)$ is the real part of the global quasimomentum 
for $(1.13)$.}

The formulas
$(2.31a),\; (2.32a)$ for the case of smooth potential are a particular case
 of results of \cite{Sh} about density of states of multidimensional
selfadjoint elliptic operators with almost periodic coefficients.

Consider now the one-dimensional Schr\"odinger equation
$$ 
-\psi^{\prime\prime} + q(x)\psi = E\psi,\ \ x\in{\Bbb R}, 
\eqno(2.33)
$$
with a potential consisting of $n$ not necessarily identical cells. 
More precisely, we suppose that
$$
{\Bbb R}=\bigcup\limits_{j=1}^nI_j,\ \ n\in{\Bbb N},
\eqno(2.34a)
$$
$$
\begin{array}{l}
I_1=] -\infty, x_1],\ I_j=[x_{j-1}, x_j]\ \ {\rm for}\ \ 1<j<n,\
I_n=[x_{n-1}, -\infty [,\\
\mathstrut\\
-\infty< x_{j-1}< x_j < +\infty\ \ {\rm for}\ \ 1< j< n,
\end{array}
\eqno(2.34b)
$$
$$
q(x)=\sum_{j=1}^nq_j(x), 
\eqno(2.34c)
$$
$$
\begin{array}{l}
q_j\in L^1({\Bbb R}),\ \ q_j={\bar q}_j,\ \ supp\,q_j\subseteq I_j\ \ 
{\rm for}\ \ 1\le j\le n,\\
\mathstrut\\
\int_{\Bbb R}(1+|x|)|q_1(x)|dx < \infty,\ \ 
\int_{\Bbb R}(1+|x|)|q_n(x)|dx < \infty.
\end{array}
\eqno(2.34d)
$$
In the present paper we obtain, in particular, the following result.

{\bf Theorem 3.}
{\it Under assumptions (2.34), the following estimate holds:
$$
|F(] -\infty, E[) - \sum_{j=1}^nF_j(] -\infty, E[)|\le n-1\ \
{\rm for}\ \ E\le 0,
\eqno(2.35)
$$
where   $F(] -\infty, E[)\ \  (F_j(] -\infty, E[)$, respectively) denotes
the distribution function of discrete spectrum for the scattering problem
for (2.33) (for the one-dimensional Schr\"odinger equation with the
potential $q_j$, respectively) on the whole line.}

In addition, for $E=0$ there is the following estimate
$$1-n+\sum_{j=1}^nF_j(] -\infty, 0[)\le F(] -\infty, 0[)\le
\sum_{j=1}^nF_j(] -\infty, 0[) \eqno(2.36)$$
given earlier in \cite{AKM2} as a development of results of \cite{K} 
and \cite{SV}.
The estimate (2.36) is more precise than (2.35) for $E=0$. However,
for fixed $E<0$ and, at least, for $n=2$ the estimate (2.35) is the best
possible, in general.

The proof of Theorem 3 is given in Section 4.

\section*{3. Auxiliary results.}

To prove {\bf Theorems 1, 2, 3} we use auxiliary results given below
separated into five parts.

\medskip
{\large\Roman{one}.}  We consider the Schr\"odinger equation
$$
-\Psi''+v(x)\Psi=E\Psi,\;\;x\in\Bbb R,
\eqno(3.1)
$$
where
$$
v=\bar v,\quad v\in L_{{\it loc}}^1({\Bbb R}),\quad E\in\Bbb R.
\eqno(3.2)
$$

Under assumptions $(3.2)$, the following formulas hold:
$$
\Bigl|\arg\bigl(\Psi (x)+\bigg.i\Psi'(x)\bigr)\biggr|_0^y-
\arg\bigl(\varphi (x)+\biggl.i\varphi'(x)\bigr)\biggr|_0^y\:\Bigr|<\pi ,
\eqno(3.3)
$$
$$
0<\frac{\pi}{2}-\arctan\,\frac{\varphi'(0)}{\varphi(0)}-
\arg\bigl(\varphi (x)+\biggl.i\varphi'(x)\bigr)\biggr|_0^y\: -
\pi N(\varphi,]0,y[)\le\pi\quad\mbox{if   }\varphi (0)\ne 0,
$$
$$
0<-\arg\bigl(\varphi (x)+\biggl.i\varphi'(x)\bigr)\biggr|_0^y\: -
\pi N(\varphi,]0,y[)\le\pi\quad\mbox{if   }\varphi (0)=0,
\eqno(3.4)
$$
for any non-zero real-valued solutions $\Psi $ and $\varphi$ of
$(3.1)$, where $\arg f(x)$ denotes an arbitrary continuously dependent
on $x$ branch of the argument of $f(x),\;\arctan r\in ]-\frac{\pi }{2},\; 
\frac{\pi }{2}[,$ for any $r\in{\Bbb R}$, $N(\varphi,]0,y[)$ denotes 
the number of zeroes of	$\varphi (x)$ in $]0,y[,\;y>0$. 

For the case of bounded potential these results were used, actually,
in Chapter 8 of \cite{CL} and in Section 4 of \cite{JM}.

\medskip
{\large\Roman{two}.} Under assumptions $(1.2),\,(1.3),\,(1.14)$, 
the following formulas hold:
$$
\Bigl|\arg\bigl(\varphi (x,E)+\biggl.i\varphi'(x,E)\bigr)\biggr|_0^{na}
\:+nap(E)\Bigr| <\pi ,\quad\mbox{for}\;\; E\in\bar\Lambda^f ,
\eqno(3.5a)
$$
$$
0\le -\pi^{-1}\arg\bigl(\varphi (x,E)+\biggl.i\varphi'(x,E)\bigr)
\biggr|_0^{na}
\:-[\pi^{-1}nap(E)]<1,\quad\mbox{for}\;\; E\in {\Bbb R}\setminus\bar\Lambda^f ,
\eqno(3.5b)
$$
for any non-zero real-valued solutions $\varphi$ of $(1.1)$, where 
one takes an arbitrary continuously dependent on $x$ branch of the
argument, $p(E)$ is the real part of the global quasimomentum 
and $\bar\Lambda^f$ is the closure of the forbidden energy set 
for $(1.13)$, $[r]$ is the integer part of $r\ge 0$.

The estimate $(3.5a)$ follows from $(3.3)$ and the formula
$$
-arg\,(\psi(x,E)+i\psi^{\prime}(x,E))\big|_0^{na}=nap(E)
$$
for $E\in {\bar\Lambda}^f$ and any non-zero Bloch solution $\psi(x,E)$ of
$(1.13)$. We obtain $(3.5b)$ using

(1) the left-hand side of inequalities $(3.4)$,

(2) Theorem 1.2 of Chapter 8 of \cite{CL},

(3) the representation of the monodromy operator $M(E)$ for $E\in{\Bbb R}
\backslash {\bar\Lambda}^f$ as the rotation matrix on the angle $ap(E)$
clockwise for an appropriate basis in the space of solutions
(identified with the space of the Cauchy data at $x=0$) to $(1.13)$ at
fixed $E$ ,

(4) the fact that the integer part $[-\pi^{-1}arg\,(\chi_1(\varphi(s),
\varphi^{\prime}(s))+i\chi_2(\varphi(s),\varphi^{\prime}(s)))\big|_0^y]$

\noindent
(where
$(\chi_1(\varphi(s),\varphi^{\prime}(s)),\chi_2(\varphi(s),
\varphi^{\prime}(s)))$ are the coordinates of the Cauchy data

\noindent
$(\varphi(s),\varphi^{\prime}(s))$ at $x=s$ of a
non-zero real-valued solution $\varphi$ of $(1.13)$ at fixed 
$E$ with respect to a fixed (independent of $s$) basis (in the 
space of the Cauchy data at $x=s$) for which the change of variables 
$(\varphi,\varphi^{\prime})\to (\chi_1,\chi_2)$
has a positive determinant) is independent of the basis.

\medskip
{\large\Roman{three}.} Under assumptions $(1.5)$, the following formula 
holds:

$$
F_{sc}(]-\infty,E[)=N(\varphi_\pm(\bullet ,E),]-\infty ,\infty [),\quad
E\le 0,
\eqno(3.6)
$$
where $F_{sc}(]-\infty,E[)$ is the number of bound states with energies
in $]-\infty,E[$ for the equation $(1.4),\;\;\varphi_\pm (x,E)$ are 
solutions of $(1.4)$ such that
$$
\begin{array}{l}
\varphi_+(x,E)=e^{-\kappa x}(1+o(1)),\ \kappa=i\sqrt{E}\ge 0,\ \ {\rm as}\ \
x\to +\infty\cr
\varphi_-(x,E)=e^{\kappa x}(1+o(1)),\ \kappa=i\sqrt{E}\ge 0,\ \ {\rm as}\ \
x\to -\infty,
\end{array}
$$

$N(\varphi(\bullet ,E),]-\infty ,\infty [)$ is the number of zeroes
of $\varphi (x,E)$ in $]-\infty ,\infty [$ (with respect to $x$).

If, in addition to (1.5), $v(x)\equiv 0$ for $x<x_1$, then
$$
0\le N(\psi_-(\bullet,E), ] -\infty, \infty [) - N(\varphi_-(\bullet,E),
] -\infty, \infty [)\le 1,\ \ E\le 0,
\eqno(3.7)
$$
where $\psi_-(x,E)$ is the solution of (1.4) such that
$$
\psi_-(x,E)=e^{-\kappa x},\ \ \kappa=i\sqrt{E}\ge 0, \ \ {\rm for}\ \
x<x_1.
$$

One can obtain $(3.6)$ generalizing the proof of the Theorem 2.1
of Chapter 8 of \cite{CL} and using properties of $\varphi (x,E)$ given
in Lemma 1 of Section 2 of \cite{DT}.

The same arguments that prove (3.3), (3.4) prove also (3.7) (taking into
account that
$$
{\pi\over 2} - \arctan{\psi_-^{\prime}(x_1,E)\over \psi_-(x_1,E)}\ge
{\pi\over 2} - \arctan{\varphi_-^{\prime}(x_1,E)\over \varphi_-(x_1,E)},
$$
where $\arctan\,r\in ] -\pi/2, \ \pi/2 [$ for $r\in{\Bbb R}$).

Remark.
For the case when $E$ is a bound state energy and, as a corollary,
$\varphi_{\pm}(x,E)$ is a bound state, the formula (3.6) was mentioned, for
example, in $\S 1$ of Chapter 1 of \cite{NMPZ}. Completing the present paper
we have found that the statement of the formula (3.6) in the general case
was given in Proposition 10.3 of \cite{AKM1}.

\medskip
{\large\Roman{four}.} Let
$$
\varphi (x,E)=ae^{\kappa x}+be^{-\kappa x}\quad\mbox{for  }x\ge y,
\eqno(3.8)
$$
where $a,b\in {\Bbb R},\;a^2+b^2\ne 0,\;\kappa >0,\;y>0$.

Then
$$
\varphi (x)\ne 0\;\mbox{for  }x\ge y\;\mbox{if }\varphi (y)>0,\;\;
\kappa\varphi (y)+\varphi'(y)\ge 0;
\eqno(3.9a)
$$
$$
\varphi (x)\;\mbox{ has a single zero for  }x\ge y\;\mbox{if }
\varphi (y)\ge 0,\;\;\kappa\varphi (y)+\varphi'(y)<0;
\eqno(3.9b)
$$
$$
\varphi (x)\ne 0\;\mbox{for  }x\ge y\;\mbox{if }\varphi (y)<0,\;\;
\kappa\varphi (y)+\varphi'(y)\le 0;
\eqno(3.9c)
$$
$$
\varphi (x)\;\mbox{ has a single zero for  }x\ge y\;\mbox{if }
\varphi (y)\le 0,\;\;\kappa\varphi (y)+\varphi'(y)>0;
\eqno(3.9d)
$$

Let
$$
\varphi (x)=a+bx\;\;\mbox{for  }x\ge y,
\eqno(3.10)
$$
where $a,b\in {\Bbb R},\;a^2+b^2\ne 0,\;y>0$.

Then
$$
\varphi (x)\ne 0\;\mbox{for  }x\ge y\;\mbox{if }\varphi (y)>0,\;\;
\varphi'(y)\ge 0;
\eqno(3.11a)
$$
$$
\varphi (x)\;\mbox{ has a single zero for  }x\ge y\;\mbox{if }
\varphi (y)\ge 0,\;\;\varphi'(y)<0;
\eqno(3.11b)
$$
$$
\varphi (x)\ne 0\;\mbox{for  }x\ge y\;\mbox{if }\varphi (y)<0,\;\;
\varphi'(y)\le 0;
\eqno(3.11c)
$$
$$
\varphi (x)\;\mbox{ has a single zero for  }x\ge y\;\mbox{if }
\varphi (y)\le 0,\;\;\varphi'(y)>0.
\eqno(3.11d)
$$

\medskip
{\large\Roman{five}.}  We consider the Schr\"odinger equation
$$
-\Psi''+v(x)\Psi=E\Psi,\;\;x\in\ [0,y],
\eqno(3.12)
$$
where
$$
v\in L^1([0,y]),\quad v=\bar v,\;\;y>0,
\eqno(3.13)
$$
with boundary conditions
$$
\begin{array}{l}
\Psi (0)\cos\alpha -\Psi'(0)\sin\alpha =0,\\
\\
\Psi (y)\cos\beta -\Psi'(y)\sin\beta =0,\;\;\alpha\in\Bbb R,\;\;
\beta\in\Bbb R.
\end{array}
\eqno(3.14)
$$
Without loss of generality we may assume
$$
0\le\alpha <\pi,\quad 0<\beta\le\pi.
\eqno(3.15)
$$

Consider the eigenvalues $E_j$ and the distribution function
$$
F(\Sigma)=\,\#\{E_j\in\Sigma\}\quad\mbox{(the number of eigenvalues
in an interval }\Sigma\subset\Bbb R)
\eqno(3.16)
$$
for the spectral problem $(3.12),\;(3.14)$.

Consider the solution $\varphi (x,E)$ of $(3.12)$ such that
$$
\varphi (0,E)=\sin\alpha,\quad\varphi'(0,E)=\cos\alpha .
\eqno(3.17)
$$
Under assumptions $(3.13),\,(3.15)$, the following formulas hold:
$$
F(]-\infty ,E])=\bigl[-\pi^{-1} \biggl. \arg(\varphi +i\varphi')
\biggr|_0^{y}\bigr]\quad\mbox{for  }\alpha =0,\;\beta =\pi ,
\eqno(3.18)
$$
$$
\bigl[-\pi^{-1} \biggl. \arg(\varphi +i\varphi')\biggr|_0^{y}\bigr]\le
F(]-\infty ,E])\le \bigl[-\pi^{-1} \biggl. \arg(\varphi +i\varphi')
\biggr|_0^{y}\bigr]+1\quad\mbox{for  }\alpha <\beta ,
\eqno(3.19)
$$
$$
\bigl[-\pi^{-1} \biggl. \arg(\varphi +i\varphi')\biggr|_0^{y}\bigr]+1\le
F(]-\infty ,E])\le \bigl[-\pi^{-1} \biggl. \arg(\varphi +i\varphi')
\biggr|_0^{y}\bigr]+2\;\mbox{for  }\alpha <\beta ,
\eqno(3.19)
$$
$$
F(]-\infty ,E])=\bigl[-\pi^{-1} \biggl. \arg(\varphi +i\varphi')
\biggr|_0^{y}\bigr]+1\quad\mbox{for  }\alpha =\beta ,
\eqno(3.21)
$$
where $[r]$ is defined by $(4.6)$.

For the case of bounded potential one can obtain these results using
the proof of Theorem 2.1 of Chapter 8 of \cite{CL}.

\section*{4. Proofs of Theorems 1, 2, 3 and Corollaries 1, 2.}

{\bf Proof of Theorem 1.} Consider the solution  $\varphi (x,E)$ 
of $(1.1)$ such that
$$
\varphi (x,E)=e^{\kappa x},\quad\kappa =i\sqrt{E}\ge 0,
\quad\mbox{for }x\le 0.
\eqno(4.1)
$$
Note that
$$
\varphi (0,E)=1,\quad\varphi'(0,E)=\kappa .
\eqno(4.2)
$$
Due to $(3.4),\;(4.2)$ the following formulas hold:
$$
\biggl. \arg(\varphi +i\varphi')\biggr|_0^{y}<\frac{\pi }{2}-\arctan\,\kappa ,
\eqno(4.3)
$$
$$
\begin{array}{l}
N(\varphi ,]0,y[)=\bigl[-\pi^{-1} \biggl. \arg(\varphi +i\varphi')
\biggr|_0^{y}\bigr]\\
\\
\mbox{if  }\quad -\frac{\pi }{2}\le \arctan\,\frac{\varphi'(y)}{\varphi (y)}
\le \arctan\,\kappa ,
\end{array}
\eqno(4.4)
$$
$$
\begin{array}{l}
N(\varphi ,]0,y[)=\bigl[-\pi^{-1} \biggl. \arg(\varphi +i\varphi')
\biggr|_0^{y}\bigr]+1\\
\\
\mbox{if  }\quad \arctan\,\kappa < \arctan\,\frac{\varphi'(y)}{\varphi (y)}
< \frac{\pi }{2} ,
\end{array}
\eqno(4.5)
$$
where $y>0$, 
$$
\begin{array}{l}
[r]\ \ \hbox{ is the integer part of }\ r \ \hbox{ for }\ r\ge 0,\\
\mathstrut \\
\lbrack r \rbrack = -1\ \ \ \ \hbox{ for  }\ \ \; -1< r < 0,
\end{array}
\eqno(4.6)
$$
$\arctan\,(\varphi'(y)/\varphi (y))=-\pi /2$ means that $\varphi (y)=0$.
Due to $(1.20),\;(3.5)$ the following formulas hold:
$$
[\pi^{-1}nap(E)]-1\le \bigl[-\pi^{-1} \biggl. \arg(\varphi +i\varphi')
\biggr|_0^{na}\bigr]\le [\pi^{-1}nap(E)]\quad\mbox{for  }E\in\bar\Lambda^f,
\eqno(4.7)
$$
$$
\bigl[-\pi^{-1} \biggl. \arg(\varphi +i\varphi')\biggr|_0^{na}\bigr]
=[\pi^{-1}nap(E)]\quad\mbox{for  }E\in\Bbb R\setminus\Lambda^f,
\eqno(4.8)
$$
where $[r]$ is defined by $(4.6)$ (we recall that $p(E)\ge 0$ for 
$E\in\Bbb R$).

Due to $(4.4),\;(4.5),\;(4.7),\;(4.8)$ the following formulas hold:
$$
[\pi^{-1}nap(E)]-1\le N(\varphi ,]0,na[)\le [\pi^{-1}nap(E)]+1
\quad\mbox{for  }E\in\bar\Lambda^f,
\eqno(4.9)
$$
$$
[\pi^{-1}nap(E)]\le N(\varphi ,]0,na[)\le [\pi^{-1}nap(E)]+1
\quad\mbox{for  }E\in\Bbb R\setminus\bar\Lambda^f,
\eqno(4.10)
$$
and, in addition,
$$
\begin{array}{l}
\mbox{if  }N(\varphi ,]0,na[)=[\pi^{-1}nap(E)]+1,\\
\\
\mbox{then  }\arctan\,\kappa < \arctan\,\frac{\varphi'(na)}{\varphi (na)}
< \frac{\pi }{2} ,
\end{array}
\eqno(4.11)
$$

The function $\varphi (x,E)$ is of the form $(4.1)$ for $x\le 0$, of 
the form $(3.8)$ for $x\ge na, E<0$, and of the form $(3.10)$ for 
$x\ge na, E=0$. Thus, the function $\varphi (x,E)$ has no zeroes for 
$x\le 0$ and has at most one zero for $x\ge na$. Thus,
$$
N(\varphi ,]-\infty ,na[)=N(\varphi ,]0,na[),
\eqno(4.12a)
$$
$$
0\le N(\varphi ,]-\infty ,\infty [)-N(\varphi ,]0,na[)\le 1.
\eqno(4.12b)
$$
{From} $(4.11),\;(3.9),\;(3.11),\;(4.12)$ it follows that
$$
\begin{array}{l}
\mbox{if  }\;\;N(\varphi ,]0,na[)=[\pi^{-1}nap(E)]+1,\\
\\
\mbox{then }\;\;N(\varphi ,]-\infty ,\infty [)=N(\varphi ,]0,na[).
\end{array}
\eqno(4.13)
$$
The formulas $(2.6),\;(2.7)$ follow from $(3.6),\;(4.9),\;(4.10),\;
(4.12b),\;(4.13)$.

{\bf Proof of the Corollary 1.} Consider the energies $z_i,\;
i=-1,0,\dots ,2(\# J-1)$, such that
$$
z_{-1}=-\infty ,\quad\Lambda_j^f=]z_{2j-3},z_{2j-2}[,\quad j\in J,
\eqno(4.14)
$$
where $\# J$ is the number of forbidden zones. Due to properties of $p(E)$,
for any $j\in J$ and $n\in\Bbb N$ there is $\delta^{(n)}>0$ ($\delta^{(n)}$
depends also on $p(E)$ and $a$) such that
$$
\left[\frac{nap(E)}{\pi}\right]=n{\it l}_j,\quad {\it l}_j\in\Bbb N\cup 0,
\quad\mbox{for  }E\in\bar\Lambda_j^f\cup [z_{2j-3},z_{2j-3}+\delta^{(n)}[.
\eqno(4.15)
$$

Due to $(2.6),\;(4.15)$
$$
\begin{array}{l}
F_{sc}^{(n)}(]-\infty ,E[)\in\{n{\it l}_j-1,\,n{\it l}_j,\;n{\it l}_j+1\},
\quad {\it l}_j\ge 1,\\
\\
F_{sc}^{(n)}(]-\infty ,E[)\in\{n{\it l}_j,\;n{\it l}_j+1\},
\quad {\it l}_j=0,\\
\\
\mbox{for  }E\in \bigl(\bar\Lambda_j^f\cup [z_{2j-3},z_{2j-3}+
\delta^{(n)}[\,\bigr)\,\cap\, ]-\infty ,0].
\end{array}   
\eqno(4.16)
$$
The formula $(2.8)$ follows from $(1.20),\;(4.16)$ and the fact that 
$E_j^{(n)}<0$.

{\bf Proof of Theorem 2.} Consider the solution $\varphi (x,E)$ of $(1.1)$
such that
$$
\varphi (0,E)=\sin\alpha,\quad\varphi'(0,E)=\cos\alpha .
\eqno(4.17)
$$
Due to $(1.20),\;(3.5)$ the following formulas hold:
$$
[\pi^{-1}nap(E)]-1\le \bigl[-\pi^{-1} \biggl. \arg(\varphi +i\varphi')
\biggr|_0^{na}\bigr]\le [\pi^{-1}nap(E)]\quad\mbox{for  }E\in\bar\Lambda^f,
\eqno(4.18)
$$
$$
\bigl[-\pi^{-1} \biggl. \arg(\varphi +i\varphi')\biggr|_0^{na}\bigr]
=[\pi^{-1}nap(E)]\quad\mbox{for  }E\in\Bbb R\setminus\bar\Lambda^f,
\eqno(4.19)
$$
where $[r]$ is defined by $(4.6)$. The formulas $(2.10)$--$(2.13)$
follow from $(3.19)$--$(3.21)$, $(4.18)$, $(4.19)$.

{\bf Proof of Corollary 2.} Due to properties of $p(E)$,
for any $j\in J\setminus 1$ and $n\in\Bbb N$ there is $\varepsilon^{(n)}>0$ 
($\varepsilon^{(n)}$ depends also on $p(E)$ and $a$) such that
$$
\left[\frac{nap(z_{2j-2})}{\pi}\right]-\left[\frac{nap(z_{2j-3}-
\varepsilon )}{\pi}\right]=1
\eqno(4.20)
$$
for $0<\varepsilon\le\varepsilon^{(n)}$, where $z_i$ are the same as
in the proof of {\bf Corollary 1}.

The formula $(2.14b)$ follows from $(2.10),\;(4.20)$. The formula 
$(2.14a)$ follows from $(2.10a)$ and $(1.20)$ with $j=1$. 

Due to $(2.11),\;(4.20),\;(1.20)$, for $\alpha <\beta$,
$$
\begin{array}{l}
F^{(n)}(]-\infty ,E[)\in\{n{\it l}_j-1,\,n{\it l}_j,\;n{\it l}_j+1\},
\quad\mbox{for  }j\in J\setminus 1,\\
\\
E\in ]z_{2j-3}+\varepsilon^{(n)}[\,\cup\,\bar\Lambda_j^f,\\
\\
F^{(n)}(]-\infty ,E[)\in\{n{\it l}_j,\;n{\it l}_j+1\},\quad
\mbox{for  }j=1,\;E\in\bar\Lambda_1^f.
\end{array}   
\eqno(4.21)
$$

The formula $(2.15)$ follows from $(4.21)$.

The deduction of others formulas of {\bf Corollary 2} is similar.

{\bf Proof of Theorem 3.} Suppose, first, that $n=2$. Consider the solution
$\varphi_+(x,E)$ of (2.33) such that
$$
\varphi_+(x,E)=e^{-\kappa x}(1+o(1))\ \ {\rm as}\ \ x\to +\infty,
$$
where (here and below in this proof) $\kappa=i\sqrt{E}\ge 0$.

Note that
$$
\varphi_+(x,E)=\varphi_{+,2}(x,E)\ \ {\rm for}\ \ x\ge x_1, 
\eqno(4.22)
$$
where (here and below in this proof) $\varphi_{\pm,j}$,\ $j=1,2$, denotes
the solution of (1,4) with $v=q_j$ such that
$$
\begin{array}{l}
\varphi_{+,j}(x,E)=e^{-\kappa x}(1+o(1))\ \ {\rm as}\ \ x\to +\infty,\\
\mathstrut\\
\varphi_{-,j}(x,E)=e^{\kappa x}(1+o(1))\ \ {\rm as}\ \ x\to -\infty.
\end{array}
$$
Using (3.6) for $v=q_j$ and (4.22) we obtain that
$$
N(\varphi_+(\cdot,E), [x_1, +\infty [)\le F_2(] -\infty, E[),
\eqno(4.23)
$$
$$
N(\varphi_{-,1}(\cdot,E), ] -\infty, x_1[)\le F_1(] -\infty, E[),
\eqno(4.24)
$$
where (here and below in this proof) $N(\varphi(\cdot,E),I)$ denotes the
number of zeros of $\varphi(x,E)$ in an interval $I$ (with respect to $x$).
Using the interlacing property of zeros of solutions to (1.4) (see
$\S 1$ of Chapter 8 of \cite{CL}) we obtain that
$$
N(\varphi_+(\cdot,E), ] -\infty, x_1[)\le N(\varphi_{-,1}(\cdot,E), ] 
-\infty, x_1[)+1.
\eqno(4.25)
$$
{From} (4.23)-(4.25) it follows that
$$
N(\varphi_+(\cdot,E), ] -\infty, +\infty [)\le F_1(] -\infty, E[) +
F_2(] -\infty, E[) + 1.
\eqno(4.26)
$$
Consider now the solution $\varphi_{x_1}(x,E)$ of (2.33) such that
$$\varphi_{x_1}(x_1,E)=e^{-\kappa x_1},\ \ \varphi_{x_1}^{\prime}(x_1,E)=
-\kappa e^{-\kappa x_1}.$$
Note that
$$
\begin{array}{l}
\varphi_{x_1}(x,E)=\varphi_{+,1}(x,E)\ \ {\rm for}\ \ x\le x_1,\\
\mathstrut\\
\varphi_{x_1}(x,E)=\psi_{-,2}(x,E)\ \ {\rm for}\ \ x\ge x_1,
\end{array}
\eqno(4.27)
$$
where $\psi_{-,2}(x,E)$ is the solution of (1,4) with $v=q_2$ such that
$$
\psi_{-,2}(x,E)=e^{-\kappa x}\ \ {\rm for}\ \ x\le x_1.
$$
Using (3.6) for $v=q_j$ and (3.7) for $v=q_2$ we obtain that
$$
\begin{array}{l}
N(\varphi_{+,1}(\cdot,E), ] -\infty, x_1[)=F_1(] -\infty, E[),\\
\mathstrut\\
N(\psi_{-,2}(\cdot,E), ]x_1, +\infty [)\ge F_2(] -\infty, E[).
\end{array}
\eqno(4.28)
$$
{From} (4.27), (4.28) it follows that
$$N(\varphi_{x_1}(\cdot,E), ] -\infty, +\infty [)\ge F_1(] -\infty, E[) +
F_2(] -\infty, E[).\eqno(4.29)$$
Using the interlacing property of zeros of solutions to (1.4) we
obtain that
$$
N(\varphi_+(\cdot,E), ] -\infty, +\infty [)
\ge N(\varphi_{x_1}(\cdot,E), ] -\infty,
+\infty [) -1.
\eqno(4.30)
$$
{From} (4.29), (4.30) it follows that
$$
F_1(] -\infty, E[) + F_2(] -\infty, E[) -1\le
N(\varphi_+(\cdot,E),] -\infty, +\infty[).
\eqno(4.31)
$$
{From} (3.6), (4.26), (4.31) it follows that
$$
|F(] -\infty, E[) - \sum_{j=1}^2F_j(] -\infty, E[)|\le 1.
$$
Thus, (2.35) is proved for $n=2$.

We obtain (2.35) for the general case by induction.

The proof of Theorem 3 is completed.

Remark.
The main idea of the proof of Theorem 3 is similar to the main idea of the
short proof of (2.36) presented in \cite{AKM2} (with a reference to a referee
of \cite{AKM2}).

\end{document}